# Band structure of LaB$_6$ by an algorithm for filtering reconstructed electron-positron momentum densities

G. Kontrym-Sznajd and M. Samsel-Czekała

W. Trzebiatowski Institute of Low Temperature and Structure Research, Polish Academy of Sciences, P.O.Box 1410, 50-950 Wrocław 2, Poland, e-mail:gsznajd@int.pan.wroc.pl

M. Biasini

ENEA, via don Fiammelli 2, IT-40128 Bologna, Italy

Y. Kubo

Department of Physics, Colleges of Humanities and Sciences, Nihon University, Tokyo 156, Japan

A new method (NM) for filtering three-dimensional reconstructed densities is proposed. The algorithm is tested with simulated spectra and employed to study the electronic structure of the rare-earth compound LaB$_6$. For this system, momentum densities are reconstructed from theoretical and experimental two-dimensional angular correlation of electron-positron annihilation radiation (2D ACAR) spectra. The experimental results are in good agreement with the band structure calculated with the full-potential linearized augmented-plane-wave (FLAPW) method within the local-density approximation (LDA), apart from the detection of small electron pockets in the 15th band. It is also shown that, unlike the electron-positron enhancement, the electron-electron correlations affect noticeably the momentum density.

More details in **Phys. Rev. B 70, 125103 (2004)**

## *Main results.*

Experimental 2D ACAR spectra for LaB$_6$ were interpreted in terms of both 1D projections (to study e-e correlation effects) and reconstructed 3D densities both in the extended **p** and reduced **k** spaces. Due to the simultaneous analysis of theoretical and experimental 2D ACAR spectra and the corresponding densities, the following conclusions were drawn for LaB$_6$:

➢ The IPM describes the experimental data well and the typical momentum-dependent e-p enhancement is not observed. This effect is somewhat surprising because according to the electronic structure calculations [1,2], bands crossing the Fermi level are parabolic and one could expect some momentum-dependence of the e-p enhancement.

- The small magnitude of the anisotropy of the experimental densities points out strong anisotropic e-e correlations, in agreement with the suggestion in Ref. [2]. This is a very important result because almost all theoretical approaches devoted to the many-body effects in the e-p annihilation are based on the Carbotte and Kahana results [3] and hint that the momentum-dependent e-e correlations should not be observed in the e-p annihilation experiment.
- Electron states (both topology of FS and character of bands, seen in the positron annihilation experiment) in $LaB_6$ are well described by the FLAPW band theory. However, detailed analysis of densities in **p** space (we excluded smearing effects, experimental noise, etc.) showed that the anisotropic part of the densities clearly reflects the fact that there are necks in the 14th band. Moreover, from our analysis in **p** space we conjecture the existence of small additional FS electron pockets in the 15th band observed also in dHvA experiments [1,4]. These pockets, which sizes are smaller than the limit set by the resolution of the current 2D-ACAR spectrometers, are not detected in **k** space (there is no such possibility to reproduce remarkably high jump of densities **k** in this small region). However, the comparative analysis in **p** space between the theoretical data utilized here (where no such FS sheet occurs) and the experimental data, reconstructed adopting the NM presented in this paper, shows additional density consistent with our conjecture. Therefore, this special filter seems to be a promising tool to reveal fine features of the e-p momentum density until now precluded to the 2D ACAR experiments.

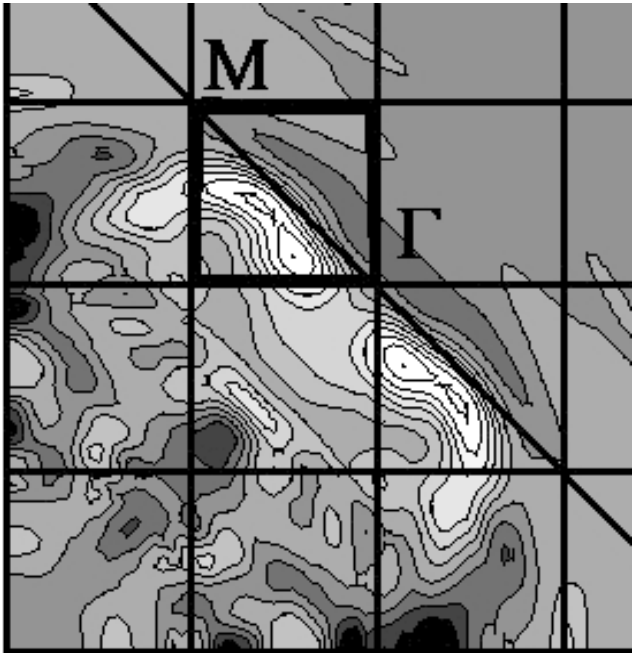

*Figure*

Anisotropic part of densities in $LaB_6$ on the plane (001), reconstructed by NM from 3 deconvoluted 2D ACAR spectra, drawn with some BZ boundaries.

It seems that we observe small electron pockets in the 15th band around the $\Sigma$ line, detected by some dHvA experiments [1,4]. It is worth noticing that the group symmetry character of the 15th band along the $\Gamma M(\Sigma)$ line is $\Sigma_4$ (see Ref. [1,2]). As such, its contribution to the momentum density in the first BZ and higher BZ's, where **p** is parallel to **k**, vanishes, as discussed by Harthorn at al. (see Ref. [5]). However, it may appear for $\mathbf{p} = \hbar(\mathbf{k} + \mathbf{G})$ with reciprocal lattice vectors $\mathbf{G}=(2\pi/a)(\xi,\eta,0)$ where $\xi \neq \eta \neq 0$. Thus, it can contribute along such ΓM lines that are parallel to the XX line from the 1st BZ. In the extended zone **p** the 14th ($\Sigma_1$) and 15th ($\Sigma_4$) bands occur in the part of the **p** space marked by the thickened square in the figure. In this area, there are the biggest contribution to the total densities ρ(**p**) coming from the 14th band (below the ΓM line) and from the 15th band (above this line). Thus, some details of the FS just in this region were able to be observed.

In the case of the theoretical calculations these elements may only be obtained if the spin-orbit interaction is included in the band structure calculations as well as the 4*f* level is displaced upward by 0.1 Ry (which may correspond to self-interactions and/or the non-local corrections to LDA - for more details see Refs. [1,2]).